# Simultaneous generation of high power, ultrafast 1D and 2D Airy beams and their frequency doubling characteristics

RAGHWINDER S. GREWAL,* ANIRBAN GHOSH, AND G. K. SAMANTA

*Photonic Sciences Lab, Physical Research Laboratory, Navarangpura, Ahmedabad 380009, Gujarat, India*
*Corresponding author: raghugrewal.singh@gmail.com*

**We report on a simple experimental scheme based on a pair of cylindrical lenses (convex and concave) of same focal length and common optical elements producing high power optical beams in 1D and/or 2D Airy intensity profiles with laser polarization as control parameter. Using an ultrafast Yb-fiber laser at 1064 nm of average power of 5 W in Gaussian spatial profile and pulse-width of ~180 fs, we have generated 1D and 2D Airy beams at an efficiency of 80% and 70%, respectively, and pulse width of ~188 fs and ~190 fs, respectively. We have measured the transverse deflection rate of 1D and 2D beams to be ~5.0 x $10^{-5}$ 1/mm and ~2.0×$10^{-5}$ 1/mm, respectively. Simply rotating the polarization state of the 1D cubic phase modulated beam in the experiment we can produce 1D and 2D Airy beams on demand. Using a 5 mm long bismuth borate ($BiB_3O_6$) we have also studied frequency-doubling characteristics of both 1D and 2D Airy beams. Like 2D Airy beam, the 1D Airy beam also produce frequency-doubled 1D Airy and an additional 1D spatial cubic structure. Like the Gaussian beams, we have observed the focusing dependent conversion efficiency for both 1D and 2D Airy beams producing green 1D and 2D Airy beams of output powers in excess of 110 mW and 150 mW for 3.4 W and 2.8 W of fundamental power respectively.**

In 1979 Berry and Balazs theoretically calculated the wave packet solution of Schrödinger equation for a free-particle in the form of Airy function [1]. This Airy packet does not spread and follows parabolic trajectory as it propagates in the absence of any external potential. The mathematical resemblance of Helmholtz equation with the Schrödinger equation has enabled experimental generation of Airy beams in the field of optics [2, 3]. Due to the peculiar properties such as self-acceleration, non-diffraction [2] and self-healing [4], the Airy beams have attracted a great deal of attention in various fields [5-7]. Typically, the optical Airy beams are generated through the Fourier transformation of a cubic phase modulated Gaussian beam. A variety of phase modulators including the spatial phase modulator (SLM) [2], cubic phase masks (CPMs) [8], ferroelectric non-linear crystals with cubic periodic gratings [9], and cylindrical lenses [10] have been used for cubic phase modulation of the Gaussian beams. While liquid-crystal-based SLMs have the intrinsic advantage of dynamic phase control through holographic technique to address any desire phase structure, however, the lower damage threshold restricts the use of SLMs for low power/energy Airy beams. On contrary, dielectric substrate based CPMs, and non-linear crystals can handle higher laser power and energy, however, these devices are wavelength specific, expensive and a single device cannot be used for both 1D and 2D cubic phase modulations.

However, the growing demand of intense and wavelength tunable Airy beam as required for various applications [11-14] has been addressed by exploiting the comatic aberration of a tilted cylindrical telescopic lens system [10]. As the cylindrical lens system imprints cubic phase modulation in 1D, therefore, the 2D cubic phase modulation requires two such systems in orthogonal orientation [10]. Although the use of cylindrical lenses make the overall experimental scheme cost effective, however, the presence of spherical aberration degrading the overall Airy beam quality demands careful system design. Therefore, the use of two such cylindrical telescopic lens systems required to generate 2D Airy beams is relatively complicated with respect to other cubic phase modulators. In addition, so far, none of the existing techniques can provide both 1D and 2D Airy beam simultaneously. Here, we report a novel experimental scheme based on a pair of concave and convex cylindrical lenses and few common optical elements available in standard optics labs generating high power 1D and 2D Airy beams on demand. Using a 5-mm-long BIBO crystal we have also studied the frequency doubling characteristics of the high power 1D and 2D Airy beams. Like 2D Airy beam [15], we have observed the appearance of additional spatial structure in second harmonic of 1D Airy beam.

The schematic of the experimental setup for simultaneous generation of 1D and 2D Airy beams is shown in Fig. 1. A 5 W

average power, Yb-fiber laser (Fianium, FP1060-5-fs), providing output pulses at 78 MHz repetition rate and spectral width of 15 nm centered at 1064 nm, is used as the primary laser. Using the intensity autocorrelator (PulseCheck, APE) we have measured the pulse-width to be ~180 fs. The laser power to the experiment is controlled using the combination of a half-wave plate ($\lambda/2$) and a polarizing beam splitter (PBS1) cube. Using two spherical lenses, L1 and L2 of focal length 50 mm and 200 mm respectively, we have expanded the laser beam of 2 mm diameter into the beam of 8 mm diameter. The experimental scheme for the simultaneous generation of 1D and 2D Airy beams comprises with PBSs (PBS2, and PBS3), cylindrical lenses (CL1 and CL2), dove prism (DP), $\lambda/2$ plate, high reflecting mirrors, M, and Fourier transforming lenses. The working principle of the experimental scheme can be understood as follows. The horizontal polarized beam in Gaussian spatial profile transmitted through the PBS2 is cubic phase modulated due to the coma aberration of the cylindrical lens systems, concave, CL1 and convex, CL2 cylindrical lens of same focal length, 50 mm. To avoid spherical and other higher order aberrations of the lens system, the lens, CL1, is tilted with respect to vertical plane such that the mechanical axis of the lens makes an angle ~35° with the beam propagation direction, whereas, the lens, CL2, is displaced in the transverse plane. The separation between the lenses is adjusted ~14 mm (longitudinally) and ~1 mm (vertical direction) according to the parameters reported in Ref. [10]. The orientation of the cylindrical lenses results 1D cubic phase modulation to the Gaussian beam in horizontal direction. The horizontal 1D cubic phase modulated (CPM) beam having horizontal polarization is transmitted though the PBS3 and guided to the dove prism (DP) using two plane mirrors, M. The DP is rotated about the longitudinal axis such that the intensity distribution of the input beam is rotated by 90°. As a result, the horizontal 1D beam transformed into vertical 1D beam after the DP (as shown in Fig. 1). The $\lambda/2$ plate is used to change the polarization state of the beam after the DP. On reflection by the mirror, M, the vertical 1D cubic phase modulated beam incidents on the PBS2. Adjusting $\lambda/2$ plate we can control the transmitted and reflected beam intensity of the PBS2. The transmitted beam of PBS2 having 1D cubic phase modulated intensity pattern is Fourier transformed using the convex cylindrical lens, CL3, of focal length 200 mm to produce 1D Airy beam. On the other hand, the vertical 1D cubic phase modulated beam reflected by PBS2 acquires cubic phase modulation in the horizontal direction while passing through the cylindrical lens combination (CL1 and CL2) and transformed into 2D cubic phase modulated beam. The vertical polarized 2D cubic phase modulated beam reflected by the PBS3 is Fourier transformed into 2D Airy beam using a plano convex lens, L3, of focal length 300 mm.

A 5-mm long and 4 mm x 8 mm in aperture BIBO crystal (C) cut for type-I ($e + e \rightarrow o$) second harmonic generation (SHG) in optical yz-plane ($\varphi = 90°$) with internal angle of $\theta = 168.5°$ at normal incidence [16], is used to study the frequency-doubling of 1D and 2D Airy beams at 1064 nm. The wavelength separator, S, is used for extraction of SHG beam from the fundamental radiation. The lenses, L4 and L5 of focal lengths, $f_4 = f_5 = f = 100$ mm in $4f$ configuration, image the 1D and 2D Airy beams to the CCD plane. The $\lambda/2$ plates placed before the lenses, CL3 and L3, are used to adjust the polarization of the fundamental beam for efficient nonlinear frequency conversion of the crystal, C. Since the polarization of the beam before the PBS2 determines the power of the transmitted and reflected beams of PBS2, adjusting the optic axis of the $\lambda/2$ plate we can generate 1D Airy beam and/or 2D Airy beam using a single experimental setup.

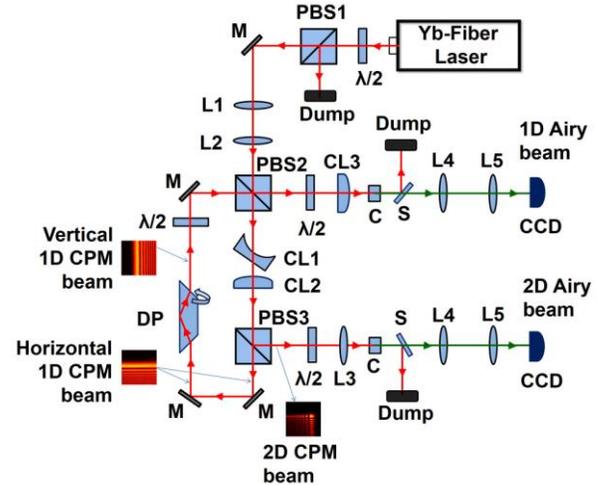

Fig. 1. Schematic of the experimental setup producing 1D and 2D Airy beams. $\lambda/2$, half-wave plate; PBS 1-3, polarizing beam-splitter cube; L1-5, plano convex lens; CL1-3, cylindrical lenses; M, mirrors; DP, Dove prism, C; BIBO crystal, S; harmonic separator, CCD1-2; CCD camera.

To generate 1D and 2D Airy beams simultaneously from the single experimental setup, we have adjusted the polarization of the 1D cubic phase modulated beam after the dove prism to 45° so that the transmitted and reflected beams of the PBS2 have equal powers. The intensity patterns of 1D and 2D beams are recorded along their propagation direction starting from the Fourier plane of CL3 and L3, respectively. The results are shown in Fig. 2. As evident from the first row of Fig. 2(a), the central lobe of the 1D beam on propagation from z = 0 (Fourier plane) to z = 18 cm shifts in the transverse plane by ~1.8 mm away from the propagation axis. Similarly, the central lobe of the 2D beam, as shown in the second row of Fig. 2(b), has a transverse shift (along vertical direction) of ~1.4 mm with propagation from the Fourier plane, z = 0, to z = 30 mm. Such transverse shift in both 1D and 2D beams with propagation confirms the beam acceleration, an important characteristics of the Airy beams. However, to get further insight, we have measured the transverse shift of both the beams with propagation at an interval of 1 cm. The results are shown in Fig. 2(b). Fitting a quadratic function (solid line) to the transverse shift of both 1D beam (solid circles) and 2D beam (solid square), we have calculated the deflection rate of 1D and 2D beams to be ~5.0 x $10^{-5}$ 1/mm and ~2.0×$10^{-5}$ 1/mm, respectively. Using the transverse intensity profiles of both 1D and 2D beams, we have measured the width (full width of half maximum, FWHM) of main lobe of respective beams along propagation distances. The width (FWHM) of 1D and 2D Airy beams are measured to be ~87 ± 4.5 µm and ~124 µm ± 4.5 µm, almost constant (within one-pixel size, 4.5 µm) over a propagation distance of 10 cm and 23 cm, respectively. Such observation clearly confirms the non-divergence property of both 1D and 2D beams. However, the main lobe width of both the beams decreases with further propagation and finally the Airy beams transform into Gaussian intensity distribution. To verify the self-healing property, we have blocked the main lobe of both 1D and 2D beams at the Fourier plane, z = 0, using a knife edge and recoded the transverse intensity distribution of the beam along propagation. The results are shown Fig. 2(c) and 2(d). As evident from Fig. 2(c), a portion of the main lobe of 1D beam is blocked at z = 1 cm, however, with propagation the beam shows sign of healing at a distance ~5 cm with complete regeneration at a distance of ~8 cm. Similarly, as evident from Fig. 2(d), the 2D beam has no central lobe at z = 1 cm, however, the beam regained its spatial intensity distribution at a propagation distance of z = 14 cm. From the self-acceleration, non-diffraction and self-healing studies we confirm the simultaneous generation of 1D and 2D Airy beams from a single experimental setup. Varying the input Gaussian beam power, we have observed a linear variation in the output power of 1D and 2D Airy beams with slope efficiency of 80% and 70%, respectively. Such high conversion efficiency along with high damage threshold of the optical elements makes the current experimental scheme most suitable for the generation of high power 1D and 2D Airy beams required to study nonlinear effects.

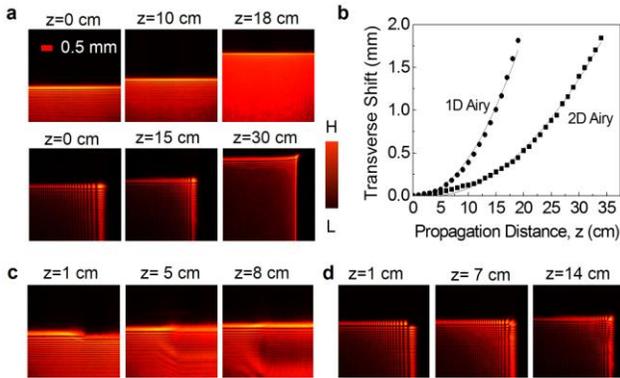

Fig. 2. (a) Transverse intensity distribution of fundamental 1D and 2D Airy beams recorded at different propagation distances. (b) Experimental (dots) trajectory of 1D and 2D Airy beam along with theoretical fit (solid line). Intensity distribution of (c) 1D and (d) 2D Airy beams recorded along beam propagation showing self-healing.

After successful generation and confirmation of high power 1D and 2D Airy beams, we have studied their nonlinear frequency-doubling characteristics. Pumping the BIBO crystals with fundamental 1D and 2D Airy beams of power 0.5 W and imaging the crystal plane (Fourier plane) at a far distance using a pair of plano convex lenses, L4, and L5, we have recorded the intensity profile of the SH Airy beams. The results are shown in Fig. 3. As evident from first row of Fig. 3(a) the green beam has an intensity distribution resembling 2D Airy beam at z = 0 (image plane), however, with propagation we observe the appearance of various mode structures similar to the previous report [15]. At z = 19 cm, we clearly observe the SH beam to carry a 2D Airy beam, two orthogonal 1D Airy beams, and an additional semi-Gaussian beam. Due to varying propagation dynamics, all those modes propagate in different directions with different acceleration as shown in Ref. [15], thus observed at a propagation distance away from z = 0 plane. We have measured the transverse shift of the 2D Airy beam to 1.7 mm in the vertical direction for beam propagation of z = 32 cm. With further propagation, the entire green beam transforms into a Gaussian beam at far field. Similarly, as evident from second row of Fig. 3(a), the frequency doubling of 1D Airy beam results in an intensity distribution of 1D Airy pattern at z = 0, with propagation to z = 13 cm, the main lobe of the beam has a transverse shift of ~0.8 mm. However, it is interesting to note that, unlike previous report [15], we observe an additional lobe structure, in the form of 1D cubic phase pattern at a propagation distance of, z = 6 cm. The additional structure has higher intensity, wider main lobe width, and transverse shift in the opposite direction of 1D Airy beam. To get further insight, we have recorded the transverse shift of both 1D Airy beam and the 1D cubic phase pattern with beam propagation at an interval of 1 cm. The results are shown in Fig. 3(b). Using quadratic and linear fit to the measured transverse shift of both 1D Airy beam (solid circles) and 1D cubic beam (solid square), respectively, we have calculated the deflection rate of 1D Airy beam and the 1D cubic beam to be ~5.5 x $10^{-5}$ 1/mm and -2.8×$10^{-3}$, respectively. The main lobe width of green 1D and 2D Airy beams are measured to be ~63 ± 4.5 μm and ~90 μm ± 4.5 μm, respectively. Like the fundamental beam, we have also measured the decrease in the main lobe width of the frequency-doubled 1D Airy beam with propagation.

We have also verified the self-healing property of the frequency-doubled 1D and 2D Airy beam by blocking the main lobe at the Fourier plane, z = 0 and measuring the intensity profile of the beam along propagation direction. The results are shown in Fig. 3(c) and 3(d). As evident from Fig. 3(c), the frequency-doubled 2D Airy beam does not have any central lobe at z = 1 cm, however, like the fundamental 2D Airy beam, the SH beam regains its spatial structure as it propagates to z = 12 cm and z = 19 cm. Similarly, as shown in Fig. 3(d), the 1D Airy beam at z = 1 cm does not carry any main lobe at the blocked region, however, starts regaining the 1D Airy beam pattern at z = 6 cm and a full recovery at z = 10 cm. It is interesting to note that unlike additional structures of 2D Airy beam (Fig. 3(c)), the 1D cubic phase structure does not reappear after the obstacle i.e., does not show self-healing property.

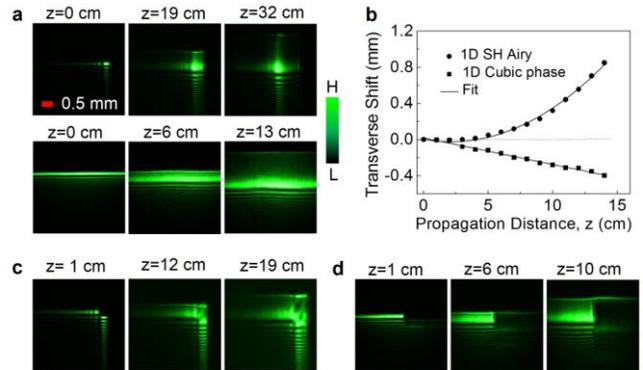

Fig. 3. (a) Transverse intensity distribution of frequency-doubled 2D and 1D Airy beams at different propagation distances. (b) Experimental (dots) trajectory of frequency-doubled 1D Airy and 1D cubic beams along with theoretical fit (solid line). Intensity distribution of green (c) 2D and (d) 1D Airy beams showing self-healing.

We have further studied the single-pass second harmonic (SH) efficiency and SH output of both 1D and 2D Airy beams. It is known that the Airy beams have large spatial extent and the side lobes contain significant portion of the overall power of the beam. However, the exponential truncation factor of finite Airy beams controlling the number of side lobes of the beam, can be adjusted by changing the beam diameter of the Gaussian beam incident to the cubic phase modulator. On the other hand, nonlinear frequency conversion processes are intensity dependent and the nonlinear crystals have limited aperture (here, 4 mm x 8 mm). To avoid beam truncation by the crystal aperture and for efficient nonlinear process we have controlled the number of side lobes of the Airy beams by reducing the Gaussian beam diameter to the cylindrical lens system. Using Gaussian beam of diameter (FWHM) ~4 mm and imaging the 1D and 2D Airy beams using two plano-convex lenses (not shown in Fig. 1) of focal lengths, $f_6$ and $f_7$, in $2f_6$-$2f_7$ configuration with different demagnifications ($D_f = f_7/f_6$) at the centre of the BIBO crystal, we have measured the SH power and single-pass conversion efficiency as a function of the pump power. The results are shown in Fig. 4. Keeping $f_6$ = 100 mm constant, we have used $f_7$ = 100, 50 and 25 mm to demagnify in the overall Airy beam size by a factor of, $D_f$ = 1, 0.5 and 0.25, respectively. As evident from Fig. 4(a), the SH power of 1D Airy beam increases quadratic (solid lines) to the fundamental power producing a maximum SH power of 18 mW, 55 mW and 111 mW for the demagnification of, $D_f$ = 1 (square), 0.5 (open circle) and 0.25 (closed circle), respectively, for a fundamental power of 3.4 W. The SH efficiency increases linear (solid lines) to the fundamental power without any sign of saturation, as shown in Fig. 4(b), proving a maximum single-pass efficiency of 0.5%, 1.66% and 3.2% for $D_f$ = 1, 0.5 and 0.25, respectively, at a constant pump power of 3.4 W. Similarly, we have measured the SH power (Fig. 4(c)) and conversion efficiency (Fig. 4(d)) of the 2D Airy beam with fundamental power for $D_f$ = 1, 0.5 and 0.25. Like 1D Airy beam, the SH power of 2D Airy beam, (see Fig. 4(c)), increases quadratic to the pump power.

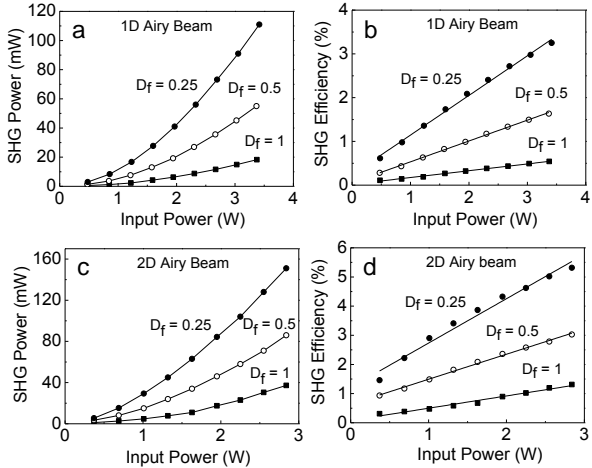

Fig. 4. Variation of (a) SHG power, (b) single-pass SHG efficiency of 1D Airy beam, and (c) SHG power, (d) single-pass SHG efficiency of 2D Airy beams for different focusing conditions.

However, for $D_f$ = 1, 0.5 and 0.25, the maximum SH power is measured to be 37 mW, 86 mW and 151 mW, respectively, at pump power of 2.8 W. The SH efficiency of the 2D Airy beam, as shown in Fig. 4(d), increases linear (solid lines) to the fundamental power without any sign of saturation, proving a maximum single-pass efficiency of 1.3%, 3% and 5.3% for $D_f$ = 1, 0.5 and 0.25, respectively, at a constant pump power of 2.8 W. However, it is to be noted that for a fixed pump power, the 2D Airy beam has higher conversion efficiency as compared to that of the 1D Airy beam. Such effect can be attributed to the effect of angular acceptance bandwidth of the BIBO crystal restricting the phase-matching of the 1D Airy beam elongated along the phase-matching direction. Using the intensity auto-correlator we have measured the pulse-width (FWHM) of the Gaussian, 1D and 2D cubic phase modulated beams at 1064 nm to be ~180 fs, 188 fs, and 190 fs, respectively. Since the Fourier transforming lenses do not significantly broaden the pulses, we can expect the Airy beams to have the same pulse-width as that of the cubic phase modulated beams. Due to the low power and beam acceleration along propagation, we could not measure the pulse-width of the green 1D and 2D Airy beams. However, we have measured the pulse-width of the SH beam to be ~197 fs generated by the Gaussian pump beam focused in 5 mm long BIBO crystal using a lens, $f$ = 150 mm. Therefore, we expect the frequency-doubled 1D and 2D Airy beams to have higher temporal width as compared that of the fundamental beams. We have measured the peak-peak power fluctuation of the green 1D and 2D Airy beams to be 6.2% and 6.4% over 1 h.

In conclusion, we have demonstrated the generation of high power 1D and 2D Airy beams on demand from a Gaussian beam using a single experimental setup at high conversion efficiency. The single-pass SHG of the Airy beam show that like 2D Airy beam, the frequency-doubling of 1D Airy beam produces 1D Airy beam and additional spatial structure at new wavelength. We have also observed the focusing dependent conversion efficiency of both 1D and 2D Airy beams. The novel experimental scheme can also be used for ultrafast lasers with narrow pulses by replacing the dove prism with a pair of ultrafast mirrors in image rotation configuration to avoid pulse broadening.

**References**
1. M.V. Berry and N. L. Balazs, Am. J. Phys. **47**, 264 (1979).
2. G. A. Siviloglou, J. Broky, A. Dogariu, and D. N. Christodoulides, Phys. Rev. Lett. **99**, 213901 (2007).
3. G. A. Siviloglou and D. N. Christodoulides, Opt. Lett. **32**, 979 (2007).
4. J. Broky, G. Siviloglou, A. Dogariu, and D. Christodoulides, Opt. Express **16**, 12880 (2008).
5. P. Polynkin, M. Kolesik, J. V. Moloney, G. A. Siviloglou, and D. N. Christodoulides, Science **324**, 229 (2009).
6. P. Zhang, J. Prakash, Z. Zhang, M. Mills, N. Efremidis, D. Christodoulides, and Z. Chen, Opt. Lett. **36**, 2883 (2011).
7. N. Voloch-Bloch, Y. Lereah, Y. Lilach, A. Gover, and A. Arie, Nature **494**, 331 (2013).
8. A. Aadhi, N. Apurv Chaitanya, M. V. Jabir, P. Vaity, R. P. Singh and G. K. Samanta, Sci. Rep. **6**, 25245 (2016).
9. T. Ellenbogen, N. Voloch-Bloch, A. Ganany-Padowicz and A. Arie, Nature Photonics **3**, 395 (2009).
10. D. G. Papazoglou, S. Suntsov, D. Abdollahpour, and S. Tzortzakis, Phys. Rev. A **81**, 061807(R) (2010).
11. C. D'Amico, A. Houard, M. Franco, B. Prade, A. Mysyrowicz, A. Couairon, and V. T. Tikhonchuk, Phys. Rev. Lett. **98**, 235002 (2007).
12. J. M. Manceau, A. Averchi, F. Bonaretti, D. Faccio, P. Di Trapani, A. Couairon, and S. Tzortzakis, Opt. Lett. **34**, 2165 (2009).
13. S. Tzortzakis, D. Anglos, and D. Gray, Opt. Lett. **31**, 1139 (2006).
14. J. Kasparian, M. Rodriguez, G. Méjean, J. Yu, E. Salmon, H. Wille, R. Bourayou, S. Frey, Y.-B. André, A. Mysyrowicz, R. Sauerbrey, J.-P. Wolf, L. Wöste, Science **301**, 61 (2003).
15. I. Dolev, I. Kaminer, A. Shapira, M. Segev, and A. Arie, Phys. Rev. Lett. **108**, 113903 (2012).
16. N. Apurv Chaitanya, A. Aadhi, R. Singh, and G. Samanta, Opt. Lett. **39**, 5419 (2014).